\documentclass[twocolumn]{aastex62}

\newcommand{\modelone}{$Z300$} 
\newcommand{\modeltwo}{$Z900$}
\usepackage{color}

\usepackage{subfigure}
\usepackage{amsmath}

\graphicspath{{./}{figures/}}

\revised{\today}

\shorttitle{LISA WD Foreground}
\shortauthors{Breivik, Mingarelli, Larson}

\begin{document}

\title{Constraining Galactic Structure with the LISA White Dwarf Foreground}

\correspondingauthor{Katelyn Breivik}
\email{kbreivik@cita.utoronto.ca}

\author[0000-0001-5228-6598]{Katelyn Breivik}
\affil{Canadian Institute for Theoretical Astrophysics, University
of Toronto, 60 St. George Street, Toronto, Ontario, M5S 1A7, Canada}
\author[0000-0002-4307-1322]{Chiara M. F. Mingarelli}
\affil{Center for Computational Astrophysics, Flatiron Institute, 162 Fifth Ave, New York, NY, 10010, USA}
\affil{Department of Physics, University of Connecticut, 196 Auditorium Road, U-3046, Storrs, CT 06269-3046, USA}
\author[0000-0001-7559-3902]{Shane L. Larson}
\affil{Center for Interdisciplinary Exploration \& Research in Astrophysics (CIERA), Northwestern University, 1800 Sherman Ave, Evanston, Illinois 60201, USA}
\affil{Department of Physics \& Astronomy, Northwestern University, 2145 Sheridan Road, Evanston, Illinois 60208-3112, USA}

\begin{abstract}

White dwarfs comprise $95\%$ of all stellar remnants, and are thus an excellent tracer of old stellar populations in the Milky Way. Current and planned telescopes are not able to directly probe the white dwarf population in its entirety due to its inherently low luminosity. However, the Galactic population of double white dwarf binaries gives rise to a millihertz gravitational-wave foreground detectable by the Laser Interferometer Space Antenna (LISA). Here we show how characterizing the angular power of the WD foreground will enable probes of the Galactic structure in a novel way to determine whether the Galactic white dwarf population traces the spatial distribution of young, bright stars, or traces a vertically heated spatial distribution associated with Galaxy's oldest stellar populations. We do this using a binary population synthesis study that incorporates different Galactic spatial distributions for the double white dwarf population. We find that the level of anisotropy in the white dwarf foreground's angular power spectrum is dependent on the vertical scale height of the population, but show that multipole coefficients from the spherical harmonic decomposition must be considered individually because of LISA's angular resolution. Finally, we show that LISA can probe the vertical scale height of the Galactic white dwarf population with an accuracy of $300\,\rm{pc}$, using the hexadecapole moment of the WD foreground.
\end{abstract}


\section{Introduction} \label{sec:intro}
The scale heights of Galactic stellar populations are a direct probe of dynamical interactions over the age of the Milky Way. Different scale heights as a function of radius can test dark matter models (e.g. \citealt{cmo19}), the Milky Way's minor merger history (e.g. \citealt{v2008}), the strength of tidal interactions from close interactions (e.g. \citealt{Bensby2010}), or constant heating through effects from the dynamical quadrupole of the Galactic bar (e.g. \citealt{Grand2016}). Since dynamical interactions are expected to occur on Gyr timescales, old stellar populations are excellent candidates for scale height measurements which trace the dynamical evolution of the Galaxy \citep[e.g.][]{Belokurov2013}. However, these populations are dim and thus difficult to observe electromagnetically throughout the Galaxy. So far, the most precise Galactic structure measurements come from fitting data from electromagnetic surveys to Galactic population synthesis simulations, \citep[e.g.][]{Robin2003, Juric2008, McMillan2011, Gao2013, Pieres2019}. These surveys have limited fields of view and observe mostly young, bright sources due to magnitude limits. Thus they do not fully probe the structure of the dimmest, oldest stellar populations.

Double white dwarf (DWD) systems are an interesting probe of the spatial structure of Galactic populations since they are the remnants of low-mass stellar progenitors which make up $95\%$ of total population. Furthermore, the population of DWDs is necessarily old since it is born from stellar progenitors with Gyr lifetimes. Importantly, gravitational wave (GW) signals from long-lived inspiraling DWDs are not suppressed or obscured by gas, dust, or other stars in the Galaxy, as their electromagnetic counterparts may be. GWs are therefore an excellent way of observing the Galactic population of DWDs and its spatial structure.

The incoherent superposition of GW signals from the Galactic DWDs form a loud foreground in the millihertz GW frequency band, detectable the Laser Interferometer Space Antenna (LISA; e.g. \citealt{lisaWhitepaper}). 
The LISA WD foreground signal has long been considered a nuisance, with significant effort being devoted to subtracting it \citep[e.g.][]{rc17}. Indeed, it is necessary to subtract the WD foreground in order to access buried GW signals like stochastic GW backgrounds \citep{Adams2001}.

In this study, we treat the WD foreground as a loud signal which can be used to constrain the scale height of the Galactic WD population. \cite{BHB2006} and \cite{Korol2019} have previously investigated LISA's ability to constrain Galactic structure from GW observations of the DWDs.
Specifically, \cite{BHB2006} studied the 1-D shape of the power spectral density ($PSD$) of the WD foreground to try to measure the vertical scale height of the Galactic disk. Their constraints depend heavily on the estimated number density of DWDs in the Galaxy -- which is currently only well understood locally \citep{Toonen2017}.  
\cite{Korol2019} predict that individual, well-localized DWDs across the Galaxy can precisely trace the Galactic scale height. However, these well-measured DWDs originate from a small subset of the total DWD population because resolved GW measurements are biased toward more massive, shorter period, or nearby binaries \citep{Lamberts2019}. 

The WD foreground is an excellent probe of the structure of Galactic DWDs since it contains contributions from the entire population. In this study, we take a similar approach to Pulsar Timing Array (PTA)~\citep{msmv13, tmg15}, and LIGO/Virgo GW background anisotropy studies~\citep{thrane09}, and decompose the foreground on a basis of spherical harmonics to characterize the WD foreground. We find that the angular power spectra of WD populations with different spatial distributions vary, but not to a degree that is measurable by LISA. Instead, we show that it possible to constrain the scale height of the DWD population to an accuracy of $\sim300\,\rm{pc}$ using the hexadecapole moment of the multipole expansion of the LISA WD foreground. 

In Section\,\ref{sec:GWS} we review the calculation of GW signals from DWDs and we describe our simulations of the Galactic population of DWDs in Section\,\ref{sec:WDpop}. We describe how we model the WD foreground anisotropy in Section\,\ref{sec:ani} and LISA's response to the WD foreground in Section\,\ref{sec:LISA_response}. In Section\,\ref{sec:reconstruction} we detail the process to reconstruct the hexadecapole moment of the foreground and report our main results in Section\,\ref{sec:results}. We finish with a discussion in Section\,\ref{sec:conclusions}. 

\begin{figure}
    \centering
    \includegraphics{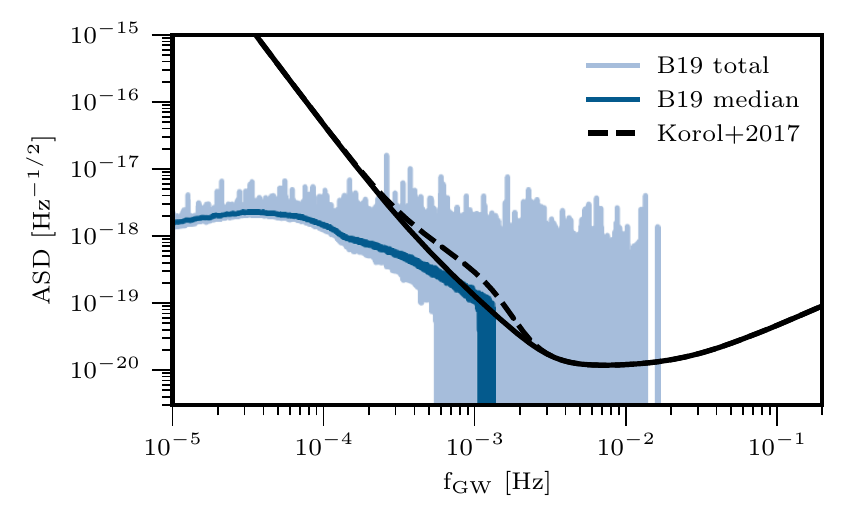}
    \caption{Amplitude spectral density vs gravitational wave frequency, $f_{\rm{GW}}$, for the entire DWD population of B19 (light blue), the running median of the population with a window of 100 frequency bins (dark blue), and the fit to the \cite{Korol2017} foreground taken from \cite{Robson2019} (dashed). The running median of B19 population is comparable to the \cite{Korol2017} foreground.}
    \label{fig:WD_foreground}
\end{figure}

\section{GWs from DWDs}
\label{sec:GWS}

Since we focus on the DWD population only, we assume all sources are circular and do not evolve due to the emission of gravitational radiation over a LISA observation time of $4$ years. We compute the polarization-averaged dimensionless strain, following \cite{Nelemans2001} as

\begin{equation}
    \label{eq:strain}
    h = 10^{-21} \Big(\frac{\mathcal{M}_c}{\rm{M_{\odot}}}\Big)^{5/3} \Big(\frac{P_{\rm{orb}}}{\rm{hr}}\Big)^{-2/3} \Big(\frac{D}{\rm{kpc}}\Big)^{-1}, 
\end{equation}
\noindent where $\mathcal{M}_c=(m_1 m_2)^{5/3}/(m_1 + m_2)$ is the chirp mass.
For stationary sources the amplitude spectral density ($ASD$) of a single DWD is 
\begin{equation}
    ASD = h\sqrt{T_{\rm{obs}}},
\end{equation}
where we assume $T_{\rm{obs}}=4\,\rm{yr}$.

To find the total GW signal from all DWDs in the Galaxy, we sum the $PSD$ of each DWD, where
\begin{equation}
    PSD = ASD^2 = h^2 T_{\rm{obs}}.
\end{equation}

\noindent The $PSD$ of the population is binned according to LISA's frequency resolution of $\Delta f = 1/T_{\rm{obs}} \simeq 8\times 10^{-9}\,\rm{Hz}$ such that the $PSD$ for each frequency bin is the sum of the $PSD$ from each DWD occupying that bin. The $ASD$ of the frequency bin is the square root of the $PSD$ of that bin. 

The process of subtracting resolved sources from the DWD population to produce an irreducible foreground is beyond the scope of this work and is discussed in several other studies \citep[e.g.][]{rc17}. Instead, we focus on the GW signal coming from the entire population which creates the WD foreground. In order to provide a comparison to previous work which has simulated a resolved source subtraction routine \citep[e.g.][]{Korol2017}, we approximate the irreducible foreground by taking a running median of the $PSD$ with a window of $100$ frequency bins. 

Figure\,\ref{fig:WD_foreground} shows a comparison of the $ASD$ of the LISA noise floor compared to the $ASD$ of the DWD population of \cite{Breivik2019}, hereafter B19. Since the orbital evolution of the DWD population is driven by GW emission, the orbital evolution scales as $\dot{f}_{\rm{orb}}\propto f_{\rm{GW}}^{11/3}$. This leads to a pileup of DWDs at lower frequencies. The foreground sharply decreases near $10\,\rm{mHz}$ because the B19 population removes all mass transferring DWDs. Generally, mass transferring DWDs are expected to occupy higher GW frequencies, and will thus not contribute to the foreground near $1\,\rm{mHz}$; see \cite{Kremer2017, Breivik2018} for a discussion of mass transferring DWDs observable by LISA. Figure\,\ref{fig:WD_foreground} also shows the irreducible foreground of \cite{Korol2017} and the running median of the B19 $ASD$. We find agreement within a factor of $\sim2-3$ between the two curves, though the B19 running median artificially cuts off the $ASD$ near $2\,\rm{mHz}$. This cutoff is a direct consequence of the truncation of the B19 foreground at higher frequencies.

\section{Simulating White Dwarf Binaries in the Milky Way}
\label{sec:WDpop}

\begin{figure*}
    \centering
    \includegraphics{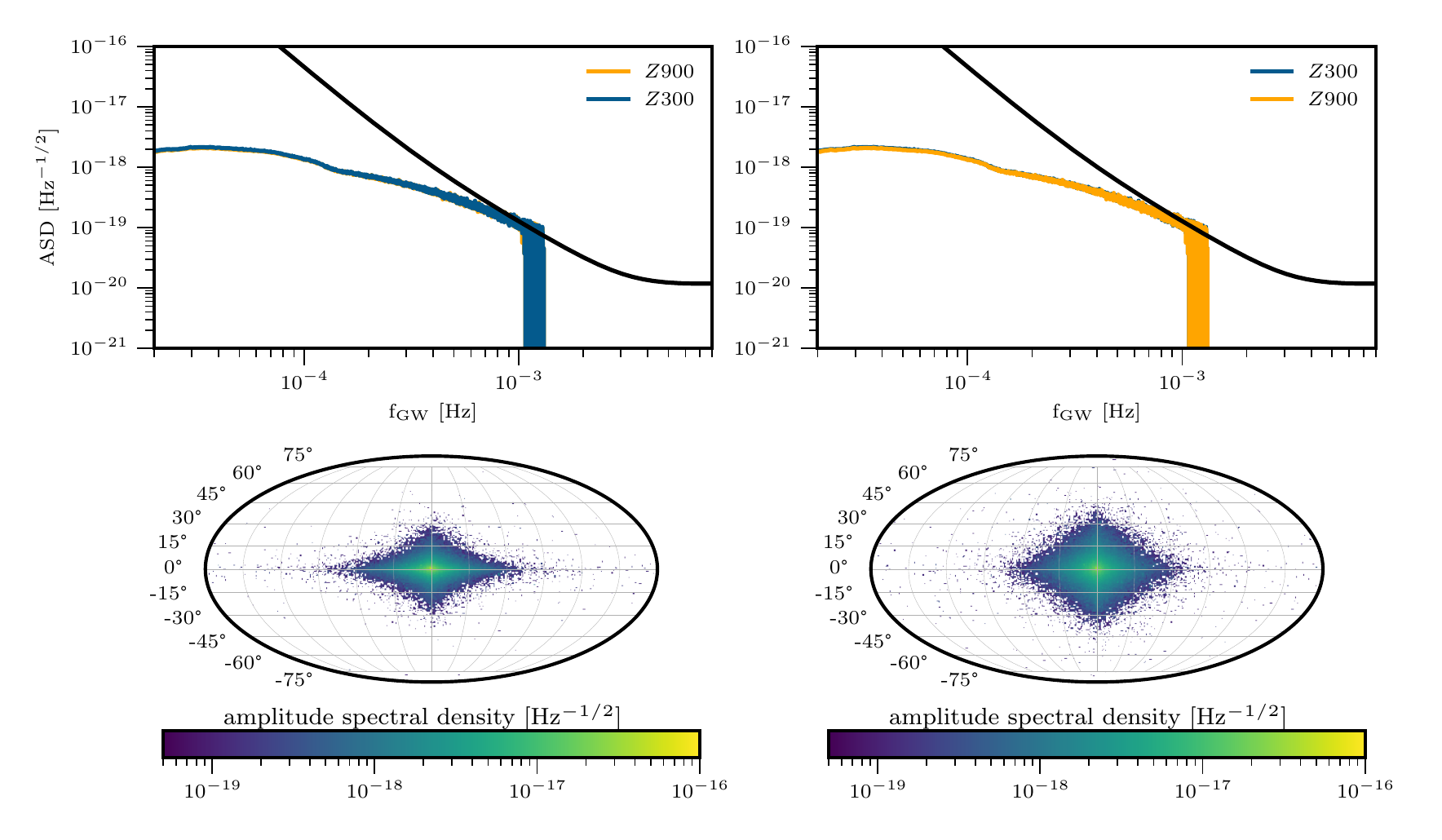}
    \caption{ The irreducible foreground 1D-amplitude spectral density (ASD; above) and the WD population (below) for \modelone\ (left, $z_h = 300\,\rm{pc}$) and \modeltwo\ (right, $z_h = 900\,\rm{pc})$. On the top row, both models are overplotted in each Figure to illustrate the nearly identical ASD. The bottom row shows the ASD in Galactocentric coordinates for all frequencies between $0.1\,\rm{mHz}$ and $10\,\rm{mHz}$. While the 2-D projection of the two models vary significantly (bottom left and right), the ASDs (top left and right) are virtually identical -- motivating our new approach. 
    }
    \label{fig:1D_v_2D}
\end{figure*}

We generate Milky Way populations of DWDs using COSMIC\footnote{cosmic-popsynth.github.io}. COSMIC is a community-developed, python-based binary population synthesis suite based on BSE \citep{Hurley2002} which includes several upgrades to binary interactions and massive star evolution, as well as models for initial binary populations and Galactic spatial distributions (see B19 for a complete discussion of these features). All DWD populations in this study use the binary evolution model described in B19. 

This study determines LISA's ability to distinguish between two models. In our fiducial model, \modelone, the Galactic DWD population traces the spatial distribution of the young, bright stellar population. In our comparison model, \modeltwo, all DWDs in the Galactic disk are distributed in a thick disk with a large vertical scale height associated with old stellar populations.

We generate DWD populations with star formation histories and spatial distributions for the thin disk, thick disk, and bulge, following the procedure detailed in B19. The thin disk is assumed to be formed from constant, solar metallicity star formation over the past $10\,\rm{Gyr}$, while the thick disk is assumed have formed from a $1\,\rm{Gyr}$ burst of uniform, $15\%$-solar metallicity star formation $11\,\rm{Gyr}$ in the past. The bulge is assumed to have formed $10\,\rm{Gyr}$ in the past with a $1\,\rm{Gyr}$ burst of uniform, solar metallicity star formation.

Following B19, the spatial distributions for \modelone\ are drawn from the models of \cite{McMillan2011}. The thin and thick disks are assumed to be azimuthally symmetric and distributed radially and vertically as

\begin{equation}
    \rho(r)\rho(z) \propto \exp{(-r/r_h)}\exp{(-z/z_h)}\, .
\end{equation}

\noindent We assume a radial scale height of $r_h=2.9\,\rm{kpc}$ and a vertical scale height of $z_h=0.3\,\rm{kpc}$ for the thin disk. The thick disk radial and vertical scale heights are  $r_h=3.31\,\rm{kpc}$ and $z_h=0.9\,\rm{kpc}$. The bulge is also assumed to be azimuthally symmetric and distributed radially and vertically as

\begin{equation}
    \rho(r') \propto \frac{\exp{[-(r/r_{\rm{cut}})^2]}}{(1+r'/r_0)^\alpha} \, ,
\end{equation}

\noindent where,

\begin{equation}
    r' = \sqrt{r^2 + (z/q)^2} \, ,
\end{equation}

\noindent and $\alpha=1.8$, $r_0 = 0.075\,\rm{kpc}$, $r_{\rm{cut}}=2.1\,\rm{kpc}$, and $q=0.5$. The mass of the thin disk is assumed to be $M_{\rm{thin}} = 4.32\times 10^{10}\,\rm{M_{\odot}}$; the mass of the thick is assumed to be $M_{\rm{thick}} = 1.44\times 10^{10}$; the mass of the bulge is assumed to be $M_{\rm{bulge}} = 8.9\times 10^9\,\rm{M_{\odot}}$ \citep{McMillan2011}. 

Our comparison model, \modeltwo, uses the same binary evolution, star formation history as \modelone. However, \modeltwo\ distributes all DWDs in the Galactic disk with vertical positions according to \modelone's thick disk scale height of $z_h=900\,\rm{pc}$. As noted above, this comparison is designed to test if the GW foreground can be used to confirm whether the DWDs trace the same Galactic distribution as the relatively younger stars observable across the Galaxy by electromagnetic surveys, or if they trace a more dynamically heated distribution associated with old stellar populations in a thick disk. 

Spatial distributions with larger vertical disk scale heights, without changes to the DWD spatial density or population number, increase the average distance to the DWD population. This, in turn, decreases the overall strength of the GW foreground since the GW signal scales inversely with distance (see \autoref{eq:strain}). Different star formation histories change the birth time of the DWD population, and thus impact the number of DWDs which radiate in the LISA frequency band  \citep[e.g.][]{Lamberts2019}. If the DWD population results from a majority of very early star formation, this leads to longer GW evolution times and thus a higher rate of WD mergers which reduces the foreground height and shift it towards higher GW frequencies. Conversely, if the DWD population is formed from primarily late star formation, the GW evolution times are shorter and result in a foreground only at lower frequencies. Different binary evolution models, particularly those pertaining to common envelope evolution, will also have a strong impact on the GW frequency distribution of the DWD population \citep[e.g.][]{Kremer2017}. We leave a thorough study of the impact of binary evolution on the angular GW power spectrum of the WD foreground as a topic of future study. 

Figure\,\ref{fig:1D_v_2D} shows the comparison between the 1-D $ASD$ running median of the WD foreground (top row) and the 2-D projection of the WD foreground's $ASD$ in Galactocentric coordinates (bottom row) from \modelone\ and \modeltwo. The differences between each model are barely distinguishable when comparing the 1-D $ASDs$, thus motivating the need for alternative approaches to measuring the vertical scale height of the DWD population with the WD foreground. The differences between each model when considering the 2-D projection of the ASD onto the sky, however, are very obvious. Distributing all disk DWDs in a thick disk results in GW signals distributed much more widely across the sky. This suggests a strong potential for inferring the WD foreground's spatial structure using LISA's discriminating power between different modes of the spherical harmonic decomposition.

\section{WD Foreground Anisotropy}
\label{sec:ani}

While \cite{BHB2006} studied the differences in the WD foreground $PSD$ as a function of vertical disk scale height, here we decompose the WD foreground on a basis of spherical harmonics to examine its angular power spectrum. This is a well-known approach to characterizing both electromagnetic and GW backgrounds. For example, anisotropy in the nanohertz GW background (accessible with PTAs) is likely generated by nearby unresolved supermassive BH binaries (SMBHBs), and/or excess GW power coming from galaxy clusters where there may be many merging SMBHBs~\citep{m17}. Various astrophysical GW signals could add incoherently and generate an anisotropic GW background signal in the LIGO/Virgo band \citep[e.g.][]{thrane09, Romano2015, jrs19, Renzini2019a, Renzini2019b}. Similar to LIGO/Virgo, several astrophysical or cosmological GW signals could produce an anisotropic GW background in the LISA band \citep{Giampieri1997,Adams2001,Cornish2001,Ungarelli2001,Kudoh2005,Taruya2005,Romano2017}, including the Galactic DWD population \citep{Seto2004,Conneely2019}.

We use the populations discussed in Section\,\ref{sec:WDpop} and their $PSDs$ to compute the angular power spectrum for each model. Each DWD's $PSD$ is captured in a pixel of a HEALPix sky map (we use NSIDE = 128, corresponding to $196,609$ pixels) for a given GW frequency. The total power on the sky is normalized to $4\pi$, and is decomposed as 
\begin{equation}
    P(\hat{\Omega})= \sum_{\ell,m}p_{\ell m}Y_{\ell m}\, ,
\end{equation}
where $\hat{\Omega}$ is the direction of GW propagation, and $Y_{\ell m}$ are the spherical harmonics. Note that we use $p_{\ell m}$ instead of the standard $c_{\ell m}$ to be consistent with the notation in \cite{Taruya2005}, hereafter TK05. We describe the anisotropy of the foreground in terms of the angular power spectrum
\begin{equation}
C_{\ell } = \sum_{m=-\ell}^{+\ell} \frac{|p_{\ell m}|^2}{2\ell+1}\, , 
\end{equation}
and normalize to the isotropic component, $C_0$, as in \cite{tmg15, m17}.

\begin{figure}
    \includegraphics{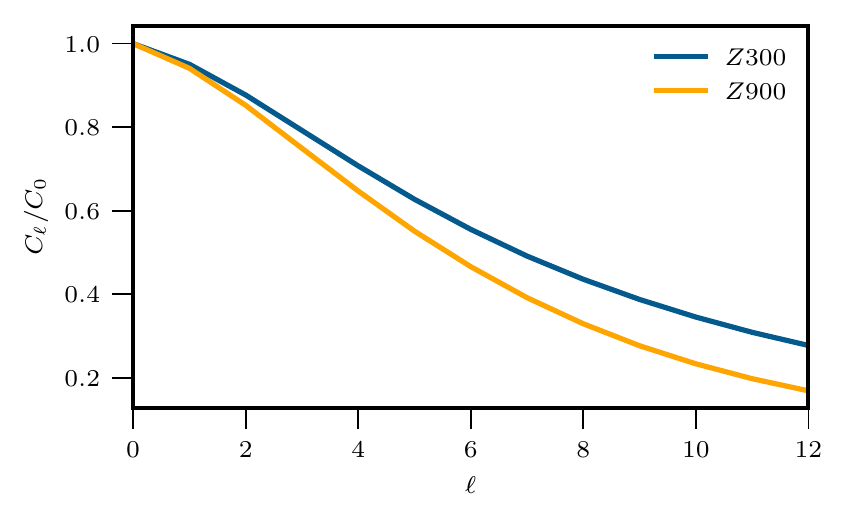}
    \caption{The angular power spectrum, $C_{\ell}$, at $f_{\rm{GW}}=3\,\rm{mHz}$ for \modelone\ (blue) and \modeltwo\ (orange). Each slice has a width of $1\,\rm{mHz}$ and is binned such that the  $f_{\rm{GW}}=3\,\rm{mHz}$ slice encompasses frequencies from $2.5\,\rm{mHz}$ to $3.5\,\rm{mHz}$. The largest differences in the power spectra occur the two models occur at multipoles $\ell \geq 8$, however LISA's angular resolution is limited to $\ell\sim4$ thus motivating the need to consider the multipole coefficients separately (see Section\,\ref{sec:LISA_response} and Section\,\ref{sec:reconstruction} for more details).} 
    \label{fig:angular_power}
\end{figure}

We show the anisotropic components of the angular power spectrum from a WD foreground frequency slice centered at $f_{\rm{GW}} = 3\,\rm{mHz}$ with a $1\,\rm{mHz}$ width created from our two models which span the possible values of vertical disk scale heights in Figure\,\ref{fig:angular_power}. We choose this frequency because it falls near LISA's minimum detector noise, but note that this method applies generally to any frequency in LISA's sensitivity band.

The level of anisotropy scales directly with the scale height of the DWD population: models with smaller vertical disk scale heights distribute WDs more anisotropically and thus produce more anisotropic GW foregrounds than models with lower disk scale heights. Indeed, \modelone\, which distributes the DWDs more closely to the plane of the Galaxy, results in a more anisotropic GW foreground with more power at higher multipoles relative to \modeltwo. The difference between the two power spectra illustrate the potential to distinguish whether the Galactic WD population is distributed similarly to observed bright stars (i.e. \modelone) or is vertically heated (i.e. \modeltwo). However, the strongest differences between the angular power spectra of the two models occur at mulitpole moments $\ell\geq8$, which is above LISA's angular resolution limit. Thus, LISA in unable to distinguish between \modelone\ and \modeltwo\ with it's current sensitivity from the angular power spectrum alone. We now consider LISA's ability to distinguish between these two models by reconstructing the multipole coefficients of each multipole moment.


\section{LISA's response to the WD foreground}
\label{sec:LISA_response}
We closely follow the methods described in TK05, which are similar to those in \cite{Seto2004}, to determine LISA's ability to observe the anisotropic WD foreground. We update LISA's noise specifications to be consistent with the current mission design \citep{Robson2019}. We use the three optimal time delay interferometry (TDI) channels: A, E, T \citep{Prince2002,Nayak2003} and consider auto- and cross-correlations for each channel. The current LISA mission design is an equilateral triangle interferometer with an arm length of $L_{\rm{arm}} = 2.5\times10^6\,\rm{km}$, which corresponds to a characteristic frequency $f_{\star}=c/(2\pi L_{\rm{arm}})\simeq19\rm{mHz}$. This characteristic frequency  defines the low frequency limit ($f_{\rm{GW}}/f_{\star} < 1$) where the formalism of TK05 is valid. Note that, our frequency slice at $3\,\rm{mHz}$ is well within the low frequency limit. 

\begin{figure*}
    \includegraphics{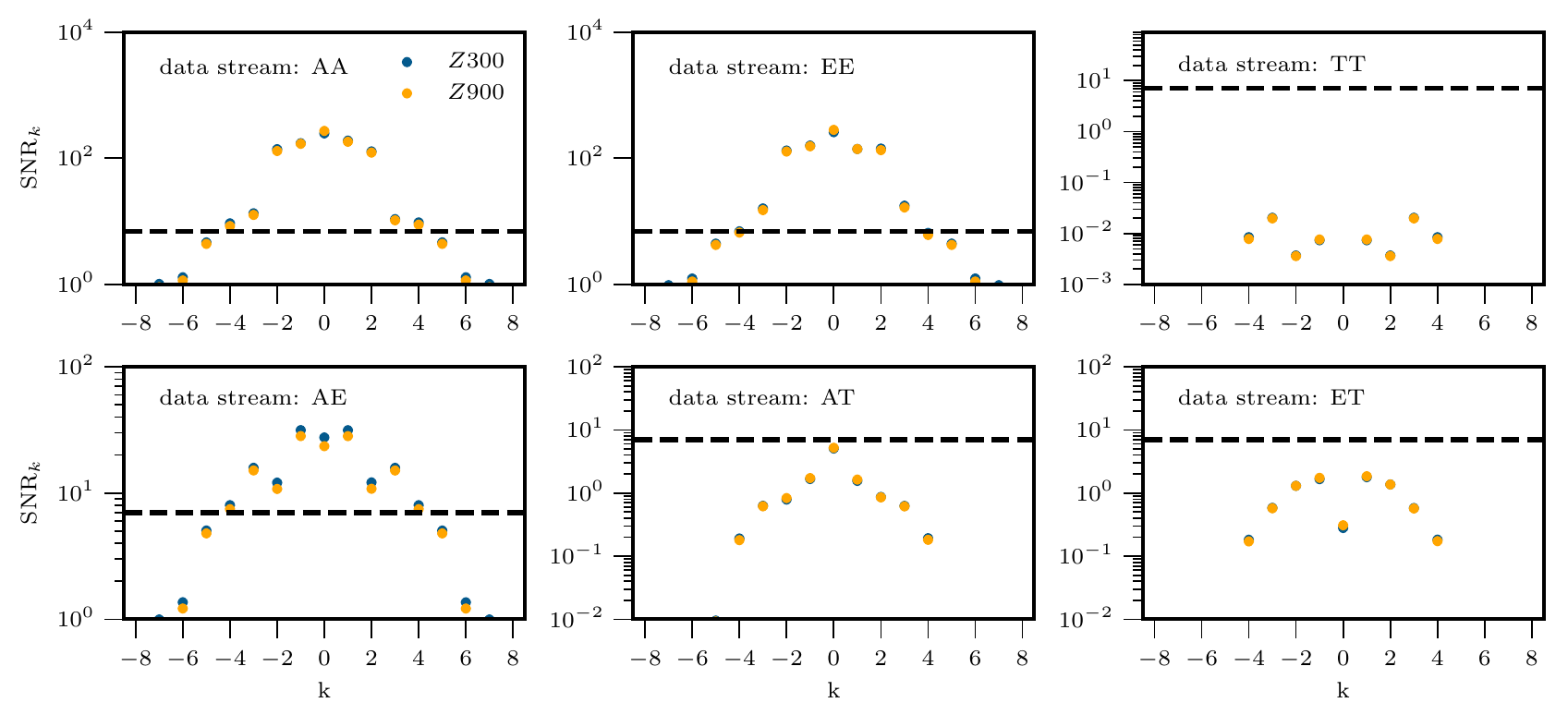}
    \caption{ $SNR_k$ for k components between -8 and 8 for each set of self and cross correlated signals for the three optimal TDI channels: A, E, and T. The colored points show $SNR_k$ at $f_{\rm{GW}} = 1\,\rm{mHz}$ for \modelone\ (blue) and \modeltwo\ (orange). The dashed black lines show where $SNR_k = 7$, which serves as a qualitative detectability criteria. Correlations AA, EE, and AE have observable signals with $|k| \leq 4$. All correlations which use the T channel have a severely reduced $SNR_k$. This is because the T channel is a `null' channel which is effectively insensitive to GWs.
    }
    \label{fig:SNR_k_1}
\end{figure*}

The correlated data streams from two channels, I and J, are given as 

\begin{equation}
    \label{eq:datastream}
    \tilde{C}_{\rm{IJ}}(t,f) = \int \frac{d\Omega}{4\pi} S_{\rm{h}}(|f|, \Omega) \mathrm{F}_{\rm{IJ}}(f,\Omega;t),
\end{equation}
where $S_{\rm{h}}(|f|, \Omega)$ is the luminosity distribution of the WD foreground
\begin{equation}
\label{eq:S_h}
    S_{\rm{h}}(|f|, \Omega) = H(f)P(\Omega) = H(f)\sum_{\ell,m}p_{\ell m}^{\ast}Y_{\ell m}^{\ast}\ ,
\end{equation}
\noindent and $\mathrm{F}_{\rm{IJ}}(f,\Omega;t)$ is LISA's time-dependent antenna pattern 
\begin{equation}
    \mathrm{F}_{\rm{IJ}}(f,\Omega;t) = \sum_{\ell,m} a_{\ell m,\rm{IJ}} Y_{\ell m}\ .
\end{equation}
\noindent In \autoref{eq:S_h}, $H(f)$ is the amplitude of the foreground, and is separable from the spherical harmonic decomposition if we consider a narrow range of frequency, in this case, a $1\,\rm{mHz}$ slice centered on $f_{\rm{GW}}=3\,\rm{mHz}$. The properties of spherical harmonics imply that $p_{\ell m}^{\ast} = (-1)^m c_{\ell, -m}$ and $a_{\ell m}^{\ast} = (-1)^{\ell -m}a_{\ell, -m}$. 

The Fourier signal of the correlated data stream is then described in terms of the spherical harmonic decomposition of the WD foreground and antenna pattern as
\begin{equation}
    \label{eq:fourier_datastream}
    \tilde{C}_{\rm{IJ}}(t,f) = H(f)\frac{1}{4\pi} \sum_{\ell,m}p_{\ell m}^{\ast}a_{\ell m;\rm{IJ}}.
\end{equation}
\noindent Note that this expression excludes terms for the detector noise. In the frequency range we consider, the WD foreground signal is at least two orders of magnitude larger than the detector noise which justifies this approximation. 

The multipole coefficients $a_{\ell m;\rm{IJ}}$ from Equation\,\ref{eq:fourier_datastream} are defined in the rest frame of the sky, thus the multipole coefficients in the rest frame of the LISA detector must be transformed using the rotation matrix \citep{Allen1997,Cornish2001,msmv13}
\begin{equation}
    \label{eq:rotation}
    a_{\ell m;\rm{IJ}}(f,t) = \sum_{n=-\ell}^{\ell}e^{-i n \alpha}d_{nm}^{\ell}(\beta)e^{-i m \gamma}a_{\ell n; \rm{IJ}}^{\rm{det}}(f),
\end{equation}
\noindent where $d_{nm}^{\ell}(\beta)$ are the Wigner small d-matrices. Following TK05, for LISA's orbit we assume $\alpha=-\omega t$, $\beta = -\pi /3$, $\gamma = \omega t$, where $\omega = 2\pi /T_0$ is LISA's orbital frequency where $T_0 = 1\,\rm{yr}$. We also adopt the multipole coefficients of LISA's antenna pattern in the detector rest frame, $a_{\ell n,\rm{IJ}}^{\rm{det}}(f)$, given in Appendix A of TK05. 

Since LISA's motion is periodic, we can analyze the Fourier component of the correlated Fourier signals, $\tilde{C}_{\rm{IJ}}(t,f)$ as

\begin{equation}
\label{eq:C_k}
   \begin{split}
       \tilde{C}_{k}(f) & = \frac{1}{T_0}\int_0^{T_0}\,dt e^{-ik\omega t}     \tilde{C}_{\rm{IJ}}(t,f)\\
                        & = \frac{1}{4\pi} \sum_{\ell=0}^{\infty}\sum_{m=-\ell}^{\ell-k}p_{\ell m}^{\ast}d_{(m+k),m}^{\ell}a_{\ell, (m+k); \rm{IJ}}(f).
   \end{split}
\end{equation}
\noindent The signal-to-noise-ratio (SNR) for each Fourier component, $\tilde{C}_{k}$, is then expressed as

\begin{equation}
\label{eq:SNR_k}
   SNR_k = \sqrt{(2\Delta f T)}\frac{|\tilde{C}_{k}|}{N_k},
\end{equation}
\noindent where $\Delta f$ is the frequency bandwidth, $T$ is the observation time, and $N_k$ is the noise contribution. We assume $\Delta f = 1\,\rm{mHz}$, $T=4\,\rm{yr}$, and the noise contribution is defined as 

\begin{equation}
\label{eq:N_0}
   N_k = \sqrt{(2\Delta f T) S_n^{\rm{IJ}}},
\end{equation}
\noindent for the $k=0$ component of the self correlated signals and as

\begin{equation}
\label{eq:N_k}
   N_k = \sqrt{\rm{max}(\tilde{C}_{\rm{II,0}}\tilde{C}_{\rm{JJ,0}}, \tilde{C}_{\rm{II,0}}S_n^{\rm{JJ}}, \tilde{C}_{\rm{JJ,0}}S_n^{\rm{II}}, S_n^{\rm{II}}S_n^{\rm{JJ}})}
\end{equation}
\noindent for the $k \neq 0$ components of the self-correlated signals and all cross-correlated signals. The $S_n^{\rm{IJ}}$ values we adopt are given in \cite{Kudoh2005} with updated optimal metrology noise and acceleration noise values consistent with the current mission design \citep{Robson2019}.

We show $SNR_k$ as a function of $k$ at $f_{\rm{GW}}=3\,\rm{mHz}$ for \modelone\ and \modeltwo\ in Figure\,\ref{fig:SNR_k_1}. Both models have very similar $SNR_k$ because the amplitude of the WD foreground at $3\,\rm{mHz}$ is nearly the same for each model (Figure\,\ref{fig:1D_v_2D}). At low frequencies covered by the WD foreground, the sensitivity of the T channel to GWs is negligible compared to the other two channels, so we only consider correlations between A and E. Based on Figure\,\ref{fig:SNR_k_1} we consider the AA, EE, and AE correlated data streams for $|k| \leq 4$ for the remainder of our analysis.

\section{Reconstructing the WD foreground}
\label{sec:reconstruction}
As noted in \cite{Cornish2001, Ungarelli2001, Seto2004}, it is not possible to fully reconstruct a map of the WD foreground for all multipole moments using only the self-correlated AA and EE data streams because many of the observed Fourier components, $\tilde{C}_{k}(f)$ are functions of multiple mulitpole coefficients leading to an under-determined set of equations. In order to reconstruct a map of the WD foreground, we employ the perturbative method of TK05 valid in the low frequency approximation ($f_{\rm{GW}})/f_{\star} < 1$. This method expands each of the self- and cross-correlated data streams defined by Eq.\,\ref{eq:C_k} in orders of frequency so that the set of $\tilde{C}_{k}(f)$ equations for each channel and order in frequency can defined in matrix form as

\begin{equation}
\label{eq:C_k_i}
       \textbf{c}^{(i)}(f) = \textbf{A}^{(i)}(f) \, \textbf{p}^{(i)}(f),
\end{equation}
\noindent where $i$ is the frequency order.

In this formalism, the multipole coefficients can be approximately solved for using the singular value decomposition method for matrix $\textbf{A}$ detailed in TK05:

\begin{equation}
\label{eq:pi}
       \textbf{p}_{\rm{approx}}^{(i)}(f) = [\textbf{A}^{(i)}(f)]^+ \, \textbf{c}^{(i)}(f).
\end{equation}

\noindent We estimate the statstical error on the reconstructed $p_{\ell m}$'s, following the method outlined in Appendix E TK05, as
\begin{equation}
\label{eq:error}
       |\Delta\textbf{p}_{\rm{approx}}^{(i)}(f)|_j^2 = [\textbf{A}^{(i)}(f)]^+_{j,k} [\textbf{A}^{(i)}(f)]^{+\ast}_{j,k} \,\langle |\textbf{S}^{(i)}_{n,k}|^2 \rangle, 
\end{equation}
\noindent where $\langle |\textbf{S}^{(i)}_{n,k}|^2 \rangle$ is taken to be 

\begin{equation}
      \langle |\textbf{S}^{(i)}_{n,k}|^2 \rangle = \Big(\alpha \frac{\textbf{c}^{(i)}_k}{\textbf{SNR}_k}\Big).
\end{equation}

\noindent Here, $\alpha$ is designed to mimic the decrease in sensitivity from considering a higher order frequency term than the dominant $\textbf{c}^{(2)}$ signal of the AE correlation. As in TK05, we set $\alpha=1$ in all cases except for the third order contributions, $\textbf{c}^{(3)}$, from the AE correlated data streams where we set $\alpha=\hat{f}^{-1}\simeq6$.

LISA's antenna pattern is insensitive to many of the multipole coefficients for each data stream. The self-correlated data streams are insensitive to all odd $\ell$ modes and only sensitive to even modes up to $\ell=4$ for $\mathcal{O}(\hat{f}^2)$ contributions and $\ell=6$ for $\mathcal{O}(\hat{f}^4)$ contributions. The cross-correlated AE data stream is sensitive to odd $\ell$ modes up to $\ell = 5$ starting at $\mathcal{O}(\hat{f}^3)$ and even $\ell$ modes up to $\ell=4$ starting at $\mathcal{O}(\hat{f}^2)$. This poor angular sensitivity leads to a sparse $[\textbf{A}^{(i)}(f)]^+$ matrix and only allows for $p_{\ell m}$'s to be solved for directly for $\ell=3$ and $\ell=4$ when considering $\tilde{C}_{k}(f)$ values from $|k| \leq 4$ based on the $SNR_k$'s of our models (Figure\,\ref{fig:SNR_k_1}). TK05 show how all $\ell$ modes up to $\ell=5$ can be reconstructed using a least squares approximation. However, since the angular power spectra of \modelone\ and \modeltwo\ show the largest differences at $\ell \geq 4$ (Figure\,\ref{fig:angular_power}), and the full reconstruction of the $\ell=5$ mode requires a strong signal from the AT/ET correlated data, we focus on the hexadecapole moment where $\ell=4$.

\begin{figure}
    \centering
    \includegraphics{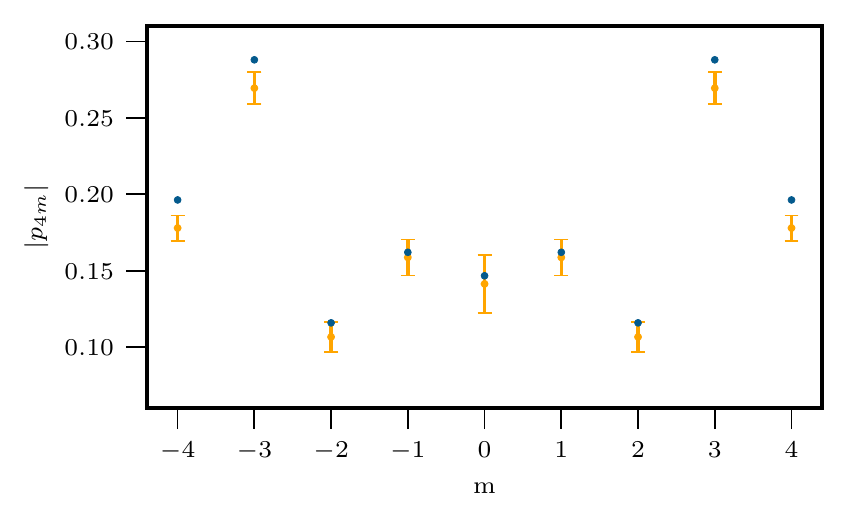}
    \caption{Reconstructed $|p_{4 m}|$ values from the AA, EE, and AE correlations for \modeltwo\ (orange) compared to the $|p_{4 m}|$ values for \modelone\ (blue). \modeltwo\ is distinguishable from \modelone\ for $|m|=3,4$, thus allowing an inference of the vertical scale height $z_h$.}
    \label{fig:plm}
\end{figure}

We plot the true and reconstructed hexadecapole moment coefficients, $|p_{4,m}|$, for \modelone\ (blue) and \modeltwo\ (orange) respectively in Figure\,\ref{fig:plm}. This choice is motivated because the scale heights for \modelone\ are observed from the bright stellar population so each multipole moment can be reconstructed based on the observed height of the WD foreground, $H(f)$, and scale heights of the thin disk, thick disk and bulge. The large statistical error of reconstructed $|p_{4,m}|$'s relative to the difference between the two models suggests that each multipole coefficient should be considered separately instead of considering the power spectrum, $C_{\ell}$. Indeed, for the $\ell=4$ mode, the reconstructed angular power spectrum for \modeltwo\ is not well constrained ($C_{4} = 0.015 \pm 0.03$). Importantly, two models are distinguishable for $|m|=3,4$ coefficients, thus allowing LISA to discriminate between \modelone\ and \modeltwo. Thus, LISA will be able to determine whether WD population traces the observed spatial distribution of the bright Galactic stellar population (\modelone), or if the WDs in the Galactic disk are vertically heated (\modeltwo).



\section{LISA's discriminating power}
\label{sec:results}

We now consider the limiting scale height at which a vertically heated disk WD population can be distingished from the spatial distribution of \modelone\ by LISA. We use the difference between the reconstructed hexadecapole moment coefficients, $|p_{4 m}|$, from different scale height models compared to hexadecapole moments resulting from the spatial distribution of \modelone\ to probe LISA's discriminating power. In particular, we consider vertical disk scale heights which fill in the intermediate values which span the scaleheight choices in \modelone\ and \modeltwo\ with $z_{\rm{h}}=350-900\,\rm{pc}$ at $50\,\rm{pc}$ intervals. This selection directly probes LISA's ability to distinguish whether the Galactic DWD population traces the spatial distribution of bright stars (e.g. \modelone) or if the DWD disk population is consistent with a larger vertical scale height typical of old stellar populations. We apply the method described above to reconstruct hexadecapole moments of the WD foreground at $f_{GW}=3\,\rm{mHz}$ for each scale height. 

\begin{figure*}
    \centering
    \includegraphics{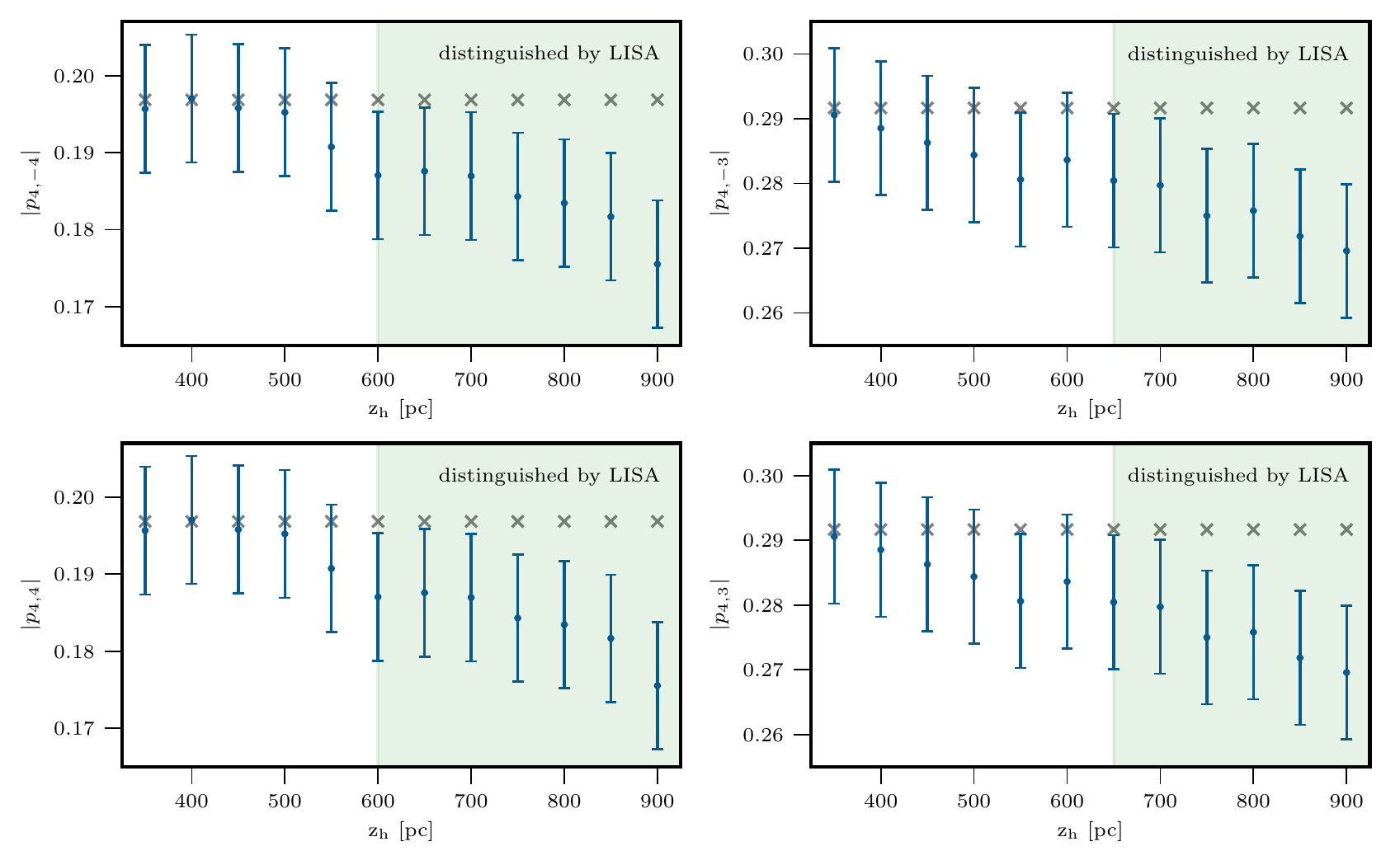}
    \caption{The reconstructed $|p_{4 m}|$ coefficients and $1\sigma$ statistical errors that are distinguishable from \modelone\ by LISA for disk vertical scale heights ranging from $z_{\rm{h}}=350-900\,\rm{pc}$ in $50\,\rm{pc}$ intervals (blue points). The multipole coefficients for \modelone\ with $z_{\rm{h}}=300\,\rm{pc}$ (gray crosses) are shown for comparison. The $|m|=3,4$ coefficients decrease with increasing scale height, becoming distinguishable from \modelone\ for models with $z_{\rm{h}}=600-650\,\rm{pc}$ as shown in the green shaded region. All other $m$ coefficients are indistinguishable from \modelone.}
    \label{fig:plm_set}
\end{figure*}

Figure\,\ref{fig:plm_set} shows the reconstructed hexadecapole moment coefficients with $|m|=3,4$ for each scale height. The multipole coefficients for \modelone\ with $z_{\rm{h}}=300\,\rm{pc}$ are shown for comparison in each panel as gray crosses. As the scale height increases, the GW power is smeared out over larger angular sizes on the sky which decreases the amplitude of the multipole coefficient. The $|m|=3,4$ coefficients are distinguishable from \modelone\ for $z_{\rm{h}}\geq550-650\,\rm{pc}$ (indicated by the green shading). It is unsurprising that the $|m|=3,4$ coefficients are constrained while the others are not are distinguished because they decompose the sky into the smallest sets of angular patches. All other hexadecapole moment coefficients are indistinguishable from \modelone\ regardless of vertical scale height, and are thus not shown. 

\begin{figure}
    \centering
    \includegraphics{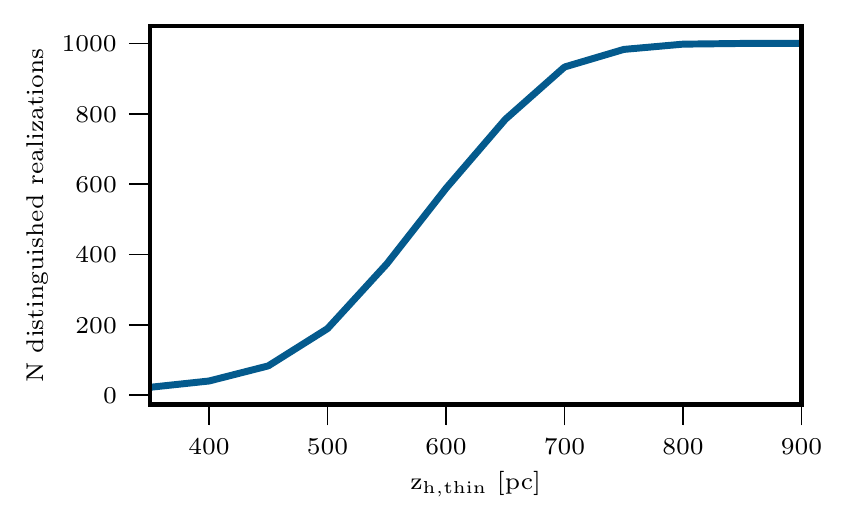}
    \caption{The number of population realizations (out of $1,000$) which have at least one multipole coefficient that is distinguishable from \modelone\ with $1\sigma$ uncertainty as a function of vertical scale height, $z_{\rm{h}}$. The fraction of distinguishable population realizations increases from $>58\%$ for $z_{\rm{h}}\geq 600\,\rm{pc}$ to $>98\%$ for $z_{\rm{h}}\geq 750\,\rm{pc}$.}
    \label{fig:distinguish}
\end{figure}

We repeat the process described above to create $1,000$ population realizations for each scale height to explore the variance in LISA's ability to distinguish between models that arises from randomly assigning the positions of the DWDs. For each realization, we deem a model distinguishable if there is at least one multipole coefficient that can be measured to be different than the multipole coefficient of \modelone. We plot the number of the realizations where at least one multipole coefficient is distinguishable at a $1\sigma$ confidence as a function of increasing vertical scale height in Figure\,\ref{fig:distinguish}.

The number of distinguishable realizations increases monotonically with increasing scale height. The trends of the multipole coefficients for each population realization are similar to those shown in Figure\,\ref{fig:plm_set}. We find that LISA will be able to distinguish whether the DWD population is distributed as $Z300$ or with scale heights $z_{\rm{h}}\geq 600\,\rm{pc}$ for $58\%$ of our realizations and with scale heights $z_{\rm{h}}\geq 750\,\rm{pc}$ for $98\%$. This corresponds to a vertical scale height measurement resolution of $300\,\rm{pc}$ and $450\,\rm{pc}$ respectively.

In comparison to our resolution of $\sim300\,\rm{pc}$, \cite{Korol2019} find that LISA can measure the scale height of the disk to an accuracy of $80\,\rm{pc}$ using observations of resolved DWDs. Both techniques provide lower resolution to electromagnetic measurements which measure the thin and thick disk scale heights to a precision of $10\,\rm{pc}-50\,\rm{pc}$ \citep[e.g.][]{Robin2003, McMillan2011, Gao2013, Pieres2019}. However, due to the inherently dim nature of WDs, electromagnetic surveys are unable to measure the WD population's structure across the Galaxy. 

The combined GW techniques, using both resolved DWDs and the WD foreground, provides a powerful probe of the spatial distribution of the WD population that is inaccessible for electromagnetic surveys. Since the two techniques are independent measures of the spatial distribution of the Galactic WD population, a consistent measurement between the two suggests that the resolved DWDs are a representative sample of the Galactic population. However, in the case where the two techniques disagree, the converse applies. In either case, constraints can be placed on spatial distribution of the Galactic population of DWDs.

\section{Discussion}
\label{sec:conclusions}
The WD foreground is a rich astrophysical GW source in the mHz frequency regime. Here, we have shown that 
analyzing the spherical harmonic decomposition of the WD foreground can be used to infer the vertical scale height of the Galactic DWD population. Measuring the spatial distribution of the DWD population separately from the birght stellar population provides insights into the stellar evolutionary history of the Galaxy. Since WDs are dim, a scale height measurement of the WD population is difficult to make with electrogmagnetic surveys; by contrast the DWD population observable by LISA is detectable over the entire volume of the Milky Way. By considering the angular power spectrum and hexadecapole moment of the WD foreground, this technique avoids both obscuration by dust, gas, and other stars as well as observational biases toward short period, high mass, and/or nearby resolved DWDs.

We used two models to illustrate the how the angular power spectrum can be used to measure the vertical disk scale height of the Galactic DWD population. \modelone\ is taken directly from the DWD population of B19 which assumes a thin disk, thick disk, and bulge population that are distributed with the same structure as the bright stellar population. \modeltwo\ assumes the same star formation history, binary evolution models, and spatial distributions for the bulge as \modelone, but vertically distributes all disk DWDs according to the spatial distribution of \modelone's thick disk. We showed that the 1-D $PSD$ of these two populations is nearly identical, but that the 2-D distribution of the $PSD$ is highly anisotropic, and significantly different between the models (Figure\,\ref{fig:1D_v_2D}). We then decomposed the GW foreground of each model on a basis of spherical harmonics to create their angular power spectra and multipole moment coefficients. 
We find that the angular power spectral shape is directly correlated with the disk's vertical scale height (Figure\,\ref{fig:angular_power}). This correlation is maintained in the reconstructed hexadecapole moment observed from the AA, EE, and AE correlated data streams (Figure\,\ref{fig:plm}). We show that the difference between the hexadecapole moments for \modelone\ and those reconstructed from different vertical scale height models can be used to infer how much scale height of the Galactic WD population deviates from the scale height of the bright stellar population (Figure\,\ref{fig:plm_set}). This presents a complementary approach to scale height measurements made through electromagnetic observations, the 1-D GW $PSD$, or DWDs resolved by LISA. 

We applied this technique to $1000$ population realizations for a set of models with scale heights varying from $z_h=350\,\rm{pc}$ to $z_h=900\,\rm{pc}$ to determine LISA's ability to resolve the vertical disk scale height of the population of Galactic DWDs (Figure\,\ref{fig:distinguish}). We find that LISA, at its current resolution limits, is capable of resolving between DWD scale height models to an accuracy of $\sim300\,\rm{pc}$ for $>50\%$ of our simulated population realizations, which is comparable to the resolution of the other measurement techniques within a factor of $6-30$. Vertical scale height measurements with greater accuracy will require angular resolutions of $\ell \geq 8$, thus motivating the demand for higher angular resolution which can be achieved through two simultaneously operating space-based GW observatories. The most straightforward way to achieve this resolution is through two space-based GW observatories operating simultaneously \citep[e.g.][]{Baker2019}. The potential for simultaneous operation of LISA and TianQin provides a distinct opportunity to gain angular resolution of the WD foreground as well as the resolved DWD population \citep{Huang2020}.

Any scale height measurements of the DWD population, which traces some of the oldest stellar populations in the Galaxy, are only possible through GW observations. The Galactic WDs are either too dim for electromagnetic surveys or obscured by gas and dust. While GW observations of resolved DWDs can measure the vertical scale height of the WD population to an accuracy of $80\,\rm{pc}$, these resolved sources could be biased towards high frequency, more massive, or nearby systems. The hexadecapole moment of the WD foreground offers a new, complementary, way to the structurre Galactic DWD population.

\section{Acknowledgements}
 The authors are grateful for helpful discussions with Ren\'{e}e Hlo\v{z}ek,  Alberto Sesana, and Kathryn Johnston and suggestions from the referee which improved the clarity of our results. K.B. acknowledges support from the Jeffery L. Bishop Fellowship. Some of the results in this paper have been derived using the healpy and HEALPix package. The Flatiron Institute is supported by the Simons Foundation. SLL acknowledges support from NASA award 80NSSC19K0323.

\software{
    Astropy \citep{Astropy2013, Astropy2018}, 
    Healpy \citep{Healpy2019}, 
    COSMIC \citep{Breivik2019, COSMIC_3.3}, 
    Matplotlib \citep{Hunter2007}, 
    Numpy \citep{Numpy2011},
    NX01 \citep{nx01},
    Pandas \citep{Pandas2010}, 
    SciPy \citep{SciPy2019}
}

\bibliography{reference}

\end{document}